\documentclass[twocolumn,aps,showpacs, prb, floatfix, reprint, superscriptaddress]{revtex4-1}
\usepackage{graphicx}
\usepackage{epstopdf}
\usepackage{longtable}

\begin{document}

\title{Molecular ordering of glycine on Cu(100): the p($2\times4$) superstructure}

\author{Zhi-Xin Hu}
\affiliation{Department of Physics, Renmin University of China,
Beijing 100872, China}

\author{Wei Ji}
\email{wji@ruc.edu.cn}

\affiliation{Department of Physics, Renmin University of China,
Beijing 100872, China}

\author{Hong Guo}
\affiliation{Centre for the Physics of Materials and Department of
Physics, McGill University, Montreal, Quebec, Canada H3A 2T8}

\begin{abstract}
Glycine molecules deposited on Cu(100) surface give rise to an
anisotropic free-electron-like (FEL) electronic dispersion in its
p(2$\times$4) superstructure, as reported in recent experiments
[Phys. Rev. Lett. {\bf 99}, 216102 (2007); J. Am. Chem. Soc. {\bf
129}, 740 (2007)]. Using density functional theory and exhaustively
calculating sixteen possible structures, we have determined the
molecular arrangement that can give the experimentally observed FEL
behavior. Eight configurations, among the sixteen, were not
investigated before in the literature and one of them (denoted
Str-3) is able to provide the FEL behavior in excellent agreement with the experiments. In addition, the particular configuration Str-3 satisfies
other criteria of the observed p(2$\times$4) superstructure, e.g.
chirality and cleavable orientation.
\end{abstract}

\received[Dated: ]{\today }
\startpage{1}
\endpage{}
\maketitle


\section{introduction}
\label{intro}

It was demonstrated that nearly free electron surface states in low-index surfaces\cite{ss} of noble metal, e.g. Cu, Ag, and Au, can be tuned by surface adsorbates\cite{kondo,corrals}, step edges\cite{steps,steps1}, artificial nanostructures\cite{rieder}, confinement along certain dimensions\cite{au-chain}, and among the others. Molecules, due to its unique capability of building various nanostructures, e.g. nano-gratings\cite{barth1,barth2} and nano-mesh\cite{cheng2010}, were employed to manipulate the surface electronic states of metals. Recently, free-electron-like (FEL) behavior was discovered on the molecular nanostructures supported by metal surfaces, rather than on the metal surfaces themselves, which has attracted great interest in both physics and chemistry communities\cite{ptcda2006,fs1,
jacs2007,prl2007,pt2007,ptcda2008,ptcda2010,zhuxy,zhaoj,feng}.
The FEL states appear to show different electron
effective masses depending on the material detail, demonstrating that
surface electronic structure can be engineered to produce desired
carrier mobility in molecular
systems\cite{ptcda2006,jacs2007,prl2007,pt2007,ptcda2008,ptcda2010}.
As an example, a FEL behavior was observed in
perylene-3,4,9,10-tetracarboxylic-3,4,9,10-dianhydride (PTCDA) molecular
islands on Ag(111)\cite{ptcda2006}, which was explained as a confined
2D interface state hybridized by the lowest unoccupied molecular state of PTCDA and the surface state of Ag underneath\cite{fs1,ptcda2008,ptcda2010}.

More recently, an {\em anisotropic} FEL behavior in a p(2x4) superstructure of glycine molecules (in the form of glycinate) on {\em isotropic} Cu(100) surface was reported\cite{jacs2007,prl2007}. This suggests a potential ability to modulate electronic structure and especially the electron effective mass of molecular monolayers or molecule-metal interfaces by using organic molecules. A key question
concerning this interesting molecular surface is why electrons move with different speeds in two different directions. In other words, why an {\it isotropic} Cu(100) surface produces an {\it anisotropic} FEL behavior after glycine molecules are deposited. Clearly, the geometric structure of the molecular system plays an essential role in generating such an anisotropic FEL behavior. However, so far the issue of geometric structure has not been adequately addressed in the literature. It is the purpose of this paper to report a systematic investigation which reveals the proper molecular configuration that supports the anisotropic FEL dispersion.

\begin{figure*}[bthp]
\includegraphics[width=14.5cm]{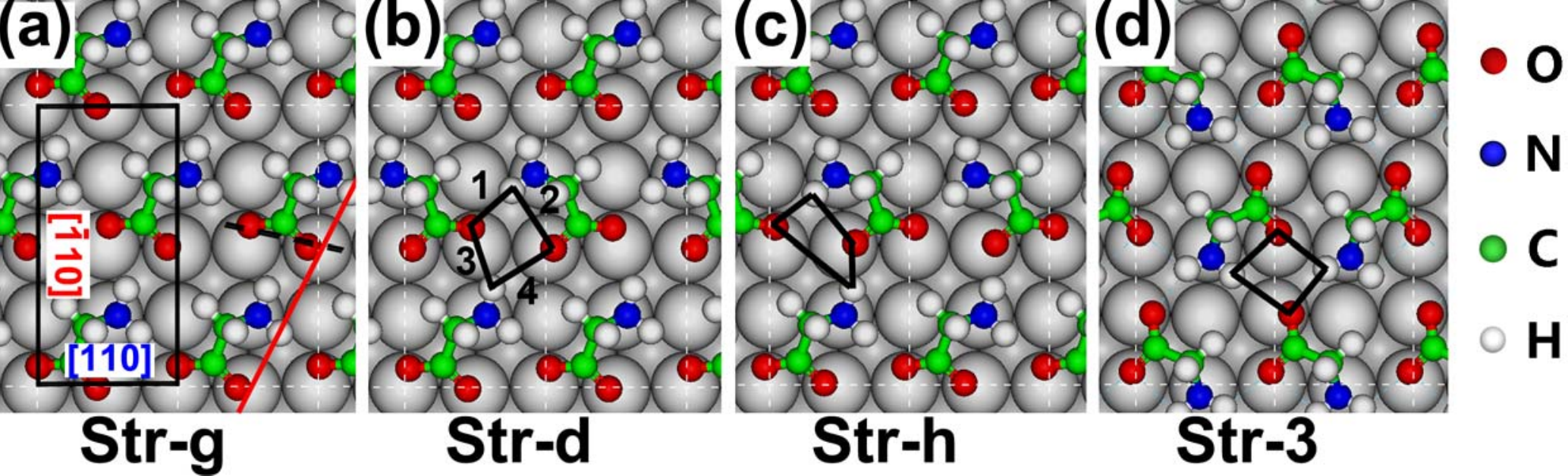}%
\caption{(Color online) Four molecular configurations. (a) to (c) are the configurations previously proposed and investigated, i.e.
c(2x4)(Str-g in this work, (a)), p(2x4) (Str-d, (b)), and p(2x4)-pg
(Str-h, (c)). (d) is a new configuration proposed in this work, in which the in-plane molecular orientation is roughly perpendicular to that in
(a) to (c). The red solid line in (a) denotes $[\overline{1}30]$, a direction included in $<3\overline{1}0>$ direction.} \label{fig:str}
\end{figure*}

About a decade ago, scanning tunneling microscopy (STM) was
employed to investigate glycine/Cu(100) by Yang~\&~Zhao's
groups\cite{yang99,yang01,yang02}. Two superstructures were found and their
atomic structures were proposed on the basis of the STM images and an
understanding of hydrogen bonds. One is a c(2$\times$4)
homochiral superstructure (denoted Str-g), as shown in Fig.\ref{fig:str}(a).
The edges of its molecular islands on the surface are along the $<3\overline{1}0>$
direction, as schematically indicated by the red solid line. Another superstructure is a p(2$\times$4) heterochiral superstructure (denoted Str-d, see Fig. \ref{fig:str}(b)), in which the edges of its molecular islands are along the $<100>$ direction. Spots in the associated low energy electron diffraction (LEED) study indicate that these two phases co-exist\cite{yang01,yang02}. Profiting from well
developed fabrication techniques, the recently observed STM images\cite{jacs2007,prl2007} clearly show that the preferred edge orientation, i.e. cleavable orientation, of p(2$\times$4) molecular islands is also along $<3\overline{1}0>$. (Here we define the ``$<3\overline{1}0>$ rule" as that the cleavable orientation of molecular islands is along $<3\overline{1}0>$.) The STM result indicates that the p(2$\times$4) structure proposed by Yang~\&~Zhao's
group, in which the molecular islands cannot be cleaved in the
$<3\overline{1}0>$ direction, appears to be {\it inconsistent} with
the recent observation\cite{prl2007}.

Photoemission diffraction (PhD) technique provides vertical atomic
information rather than local density of states as provided by STM.
In a PhD experiment\cite{phd}, a joint European group proposed
another p(2$\times$4) heterochiral superstructure, i.e. the
(2$\times$4)-pg (denoted Str-h) shown in Fig. \ref{fig:str}(c).
It was found that this structure, as a predominant structure, may
co-exist with the c(2$\times$4). Reference \onlinecite{phd} also indicates that the
(2$\times$4)-pg structure satisfies the $<3\overline{1}0>$
rule, which thus appears to be consistent with the recent STM
observation\cite{prl2007}.

Attempting to clarify the situation, Mae and Morikawa\cite{DFT} performed a density functional theory (DFT) calculation on the three structures proposed by the two
experimental groups\cite{yang99,phd}. It was found that the total energy of the proposed c(2$\times$4) (Str-g) and p(2$\times$4) (Str-d) structures are energetically very close to each other, while that of (2$\times$4)-pg (Str-h) is significantly higher. Structure (2$\times$4)-pg (Str-h) was thus deemed to be unlikely. In other words, the DFT calculation reported in Ref.\onlinecite{DFT} poses a challenge to the community that {\em none} of the proposed p(2$\times$4) structures could satisfy all previous experimental \cite{jacs2007,prl2007,yang99,yang01,yang02,phd} and theoretical\cite{DFT} results.

It was recently reported that the electronic structure of
p($2\times4$) shows an anisotropic FEL behavior as measured by scanning
tunneling spectroscopy (STS), which provides a new avenue for
solving the above discussed discrepancy. Particularly, the measured
electronic structure, i.e. the electron effective mass ($m_{e}^{*}$)
and energy levels at certain $k$-points, can be adopted as criteria
to assess all the proposed molecular configurations, finding or verifying the configuration of the observed p($2\times4$) superstructure.

\begin{figure}[bt]
\includegraphics[width=8.5cm]{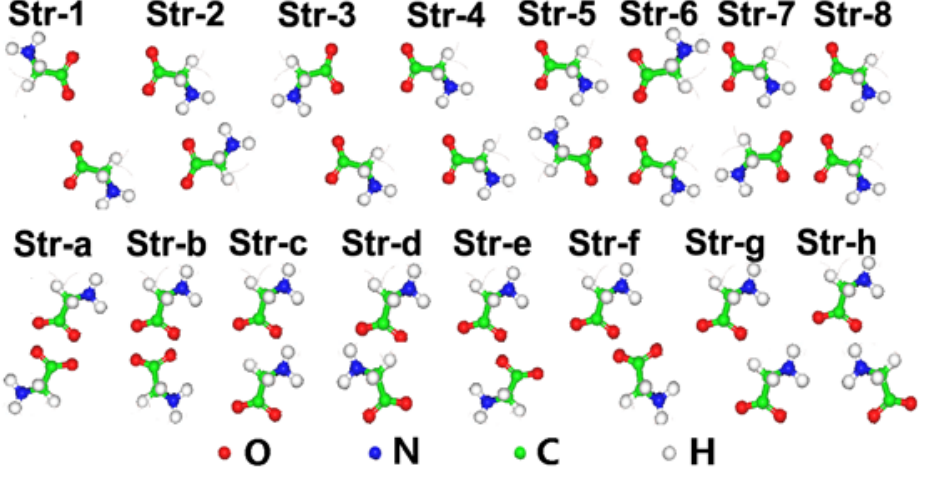}%
\caption{(Color online) According to O-O orientation, O-N direction,
the relative position of equivalent atoms, and the chirality, there
are $2^{4}=16$ configurations that should be considered, including
eight previously considered configurations (Str-a to Str-h) and
eight newly proposed configurations (Str-1 to Str-8). }
\label{fig:all-str}
\end{figure}

Based on this idea, we therefore relaxed the atomic structure and
systematically calculated the total energy of 16 configurations
including eight newly proposed ones (see Fig. \ref{fig:all-str}). For
those fully relaxed configurations that show particularly strong
stability, we calculated their electronic structures. The calculated
band structure of configuration Str-3, proposed in this work
(atomic structure shown in Fig. \ref{fig:str}(d)), is
impressively consistent with the STS results in both surface
directions. It is remarkable that the geometry of Str-3 is
highly consistent with both the $<3\overline{1}0>$ rule and the
observed chirality. Configuration Str-3 is also preferred in the
theory-experiment comparison of STM images. All these consistencies with experimental observations that other configurations cannot offer, strongly
suggest that Str-3 is most likely the configuration describing the
experimentally observed hetero-chiral p(2$\times$4) superstructure in the recent STM and STS measurements.

The rest of the paper is organized as follows. In Section II, we
present computational details including the molecular configurations
being considered and investigated. Section III presents results and
discussion. The last section provides further discussion and a
summary.

\begin{table*}
\caption{Calculated total energy and summarized information
(Chirality, cleavable orientation, and the type of lattice) of all
possible configurations.}
\begin{center}
\begin{tabular}{ccccccccccccccccc}
  \hline\hline
Str- & 1 & 2 & {\bf 3} & 4 & 5 & 6 & 7 & 8 & a & b & c & {\bf d} & e & f & {\bf g} & {\bf h} \\
\hline
Energy (eV) & 1.03 & 0.60 & {\bf 0.32} & 0.43 & 0.55 & 0.83 & 0.40 & 0.22 &  0.80 & 0.40 & 0.23 & {\bf 0.00} & 0.98 & 0.70 & {\bf 0.04} & {\bf 0.29}\\
Chirality   & N & {\bf Y} & {\bf Y} & N & N & Y & Y & N & {\bf Y} & N & N & Y & {\bf Y} & N & N & {\bf Y} \\
Orientation & Y & {\bf Y} & {\bf Y} & Y & Y & N & N & N & {\bf Y} & N & N & N & {\bf Y} & Y & Y & {\bf Y} \\
X($2\times4$) & p & p & p & c & p & p & p & - & p & p & - & p & p & p & c & p \\
\hline\hline
\end{tabular}
\end{center}
\label{tab:energy}
\end{table*}

\section{Computational details}

As discussed in Section \ref{intro}, the main issue is concerning
the p($2\times4$) superstructure. We thus only focus on this
structure in the rest of the paper. In particular, the atomic structure
of c($2\times4$) can be easily derived once the structure of
p($2\times4$) is determined. It was reported that the carboxylic
acid group (COOH) in glycine molecules is dissociated on Cu(100),
giving glycinate molecules that will be discussed in this paper.
Notice that the line jointing the two O atoms in a glycine (the
dashed line in Fig. \ref{fig:str}(a)) is roughly parallel to the
[110] direction (the shorter edge of a (2$\times$4) lattice, solid
rectangle, Fig. \ref{fig:str}(a)) in all previously proposed and
studied configurations. Nevertheless, there appears no reason why this
O-O line cannot be parallel to the $[\overline{1}10]$ direction (the
longer edge of the lattice, solid rectangle,
Fig. \ref{fig:str}(a)). Surprisingly, this possibility has not been
considered before, e.g. the configuration shown in
Fig. \ref{fig:str}(d) where the O-O line is roughly parallel to the
longer edge of the lattice. Based on this observation that
glycine may take another orientation forming a p(2$\times$4)
overlayer, 16 possible configurations were considered, in which eight
of them are proposed by this work for the first time as shown in
Fig. \ref{fig:all-str}.

All calculations were done using DFT, the generalized gradient
approximation for the exchange-correlation potential\cite{pw91}, the
projector augmented wave method\cite{paw}, and a plane-wave basis
set as implemented in the Vienna {\it ab-initio} simulation
package\cite{vasp}. Six layers of Cu atoms and a 2$\times$4
supercell, separated by a 10-layer vacuum region were employed to
model the Cu(100) surface. The molecule is only put on one side of the
slab and a dipole-correction was applied. A $k$-mesh of
$8\times4\times1$, verified by a $12\times6\times1$ one, was adopted
to sample the surface Brillouin Zone for both relaxation and total
energy calculations. The energy cut-off for the plane waves is set
to 400 eV. In structural relaxation, all atoms except for the bottom
three layers were fully relaxed until the net force on every atom is
less than 0.02 eV/\AA.

\section{Results and discussion}
\subsection{Molecular configurations}

Table \ref{tab:energy} shows the calculated relative total energies
of all configurations with respect to the lowest energy which is the
total energy of Str-d. Configurations Str-1 to Str-8 were not
considered before in the literature. Letter ``Y" (``N") in row
``Chirality" indicates that the corresponding configuration is (not)
chiral. The next row shows whether or not the configuration can be
perfectly cleaved along $<3\overline{1}0>$ directions (the
$<3\overline{1}0>$ rule). Here, term ``perfect cleaving" of a
configuration along a certain direction means that a Cu mono-atomic
step along such direction can be and only be covered by glycine
molecules in the molecular arrangement of that configuration. Row ``X($2\times4$)" shows the type of lattice of all configurations, in which Str-4 and Str-g have a c($2\times4$) lattice, the lattice of Str-8 and Str-c degrades into a ($2\times2$) one, and the others is p($2\times4$).

It was found that the relative total energies of Str-8 and Str-c are well consistent, which implies a good convergence of relative energies
between different molecular orientations to less than 0.01 eV in our
calculation. Since Str-8 and Str-c degrade into a ($2\times2$)
structure and are not observable in all experiments, they are thus
excluded from further discussion. From a purely energetic point of
view, Str-d and Str-g are the two suggested configurations which were
also preferred in the literature\cite{DFT}. The energy difference between them is 0.04 eV, consistent with the previous DFT calculation\cite{DFT}.

Purely energetic aspect aside, there are two additional factors
which may influence the determination of the observed configuration,
i.e. the chirality and the cleavable orientation. Chirality leads to
a larger entropy of a chiral structure, giving a lower free energy
at finite temperature. Although Refs. \onlinecite{yang99,yang01,yang02} did not state the preferred orientation of the island edges, it is indeed observable that most edges of glycine islands shown in the Fig. 1 of Ref. \onlinecite{jacs2007} are in the $<3\overline{1}0>$ direction. Whether or not the superstructure is cleavable along the $<3\overline{1}0>$ direction, we believe, is of importance for distinguishing the configurations.

According to Table \ref{tab:energy}, Str-d has the lowest total
energy, chirality, but is not cleavable along the $<3\overline{1}0>$
direction. The c($2\times4$) configuration Str-g has the second
lowest total energy, no chirality, but can be cleaved along the
$<3\overline{1}0>$ direction. These results are consistent with the
experimental observations of Zhao \& Yang\cite{yang99,yang01}. On
the other hand, Str-h, the PhD experiment suggested configuration, has
chirality and is cleavable along $<3\overline{1}0>$. Among those
configurations having chirality and also cleavable along $<3\overline{1}0>$,
Str-h has the lowest total energy. It also has the third lowest total energy among
all configurations (Str-8 and Str-c excluded). We thus conclude that the PhD suggested configuration is quite reasonable. Table \ref{tab:energy} shows that the total energies of all configurations proposed by this work are higher
than the previously proposed ones, i.e. Str-d, Str-g and Str-h. However, Str-3, a newly proposed configuration that has chirality and is cleavable along $<3\overline{1}0>$, has a total energy very close to Str-h (within
0.03 eV). Therefore, it should be considered as a likely candidate for
the observed p($2\times4$) superstructure.

\begin{table*}[tpb]
\caption{Structural properties obtained from the DFT calculation in
comparison to experimental data. Z$_{N(O)1}$ stands for the
difference in $z$ direction between N (O) and the first layer Cu
underneath; while Z$_{12}$ (N(O)) means that between first and
second layer Cu atoms under N(O). d$_{Cu-N(O)}$ shows the bond
length of Cu and N(O). $\theta_{N(O)}$ is the angle between the
vector perpendicular to the surface and the one formed by Cu and
N(O). There are two O atoms in each molecule, so that two values are
listed in the table.} \label{tab:str}
\begin{center}
\begin{tabular}{ccccc}
  \hline\hline
Str- & d & h & 3 & Exp.\cite{phd} \\
\hline
~~$Z_{N1}$ (\AA) ~~& ~~2.07 & 2.08 & 2.07 & 2.04$\pm$0.02\\
~~$d_{Cu-N}$ (\AA) ~~& ~~2.10 & 2.08 & 2.09 & 2.05$\pm$0.02\\
~~$Z_{12}$ (N) (\AA) ~~& ~~1.89 & 1.89 & 1.90 & 1.80$\pm$0.09 \\
~~$\theta_{N}$ ($^{\circ}$) ~~& 8.8 & 4.6 & 9.7 & 5$\pm$4\\
~~$Z_{O1}$ (\AA)~~ & ~~2.06 (2.06)~~& ~~2.05 (2.09)~~ & ~~2.03 (2.13)~~ & ~~ 2.02$\pm$0.02 ~~\\
~~$d_{Cu-O}$ (\AA)~~ & ~~2.08 (2.18)~~ & ~~2.10 (2.20)~~ & ~~2.06 (2.23) ~~ & ~~ 2.05$\pm$0.02 ~~\\
~~$Z_{12}$ (O) (\AA)~~ & ~~1.81 (1.81)~~ & ~~1.81 (1.83)~~ & ~~1.82 (1.83)~~ & ~~ 1.79$\pm$0.06 ~~\\
~~$\theta_{O}$ ($^{\circ}$)~~ & ~~7.2 (18.1)~~ & ~~12.3 (17.9)~~ & ~~10.9 (16.1)~~ & ~~ 9$\pm$2~~ \\
\hline\hline
\end{tabular}
\end{center}
\end{table*}

Recent experiments suggested that the p($2\times4$) structure should
have chirality, can be perfectly cleaved along $<3\overline{1}0>$, show an isotropic free-electron-like behavior of certain bands\cite{jacs2007,prl2007}. In addition, as a basic rule, the total energy of the structure or the kinetic energy barrier to reach it, should be as low as possible. Based on our total energy calculation, the analysis of chirality and cleavable orientation, we selected three configurations Str-d, Str-h, and Str-3, for band structure calculations as described in section C below. Reasons are the following: (1) Str-d has chirality and holds the lowest total energy among all configurations; (2) Str-h and Str-3 satisfy both rules of chirality and cleavable orientation, and have relatively low total energies.

\subsection{Atomic structures}

Structural properties that the associated experimental data are available for configurations Str-d, Str-h, and Str-3, are summarized in Tab. \ref{tab:str} in which the values for Str-d are the same as the theoretical
result in Ref.\cite{cu_surf}. Theoretical values of each configuration are consistent with the corresponding values measured by photo-diffraction\cite{phd}, hence it is difficult to assign the preferred configuration according to such experiment. The table shows that the differences in distances and angles
between these three configurations at the single molecular level are
rather small, implying that the energetic difference is likely not primarily due to the tiny structural.

The intermolecular interaction, i.e. hydrogen bonding, is therefore
involved in discussion. Each configuration contains four categories of
hydrogen bonds including three intermolecular bonds and one
intramolecular bond, as marked in Fig. \ref{fig:str}(b). The
shortest bond length in Str-d is 2.00~\AA~ which almost approaches
the shorter limit of a typical hydrogen bond, indicating a stronger
strength of this bond. Although the difference between the shortest
and longest bonds in Str-d is somewhat large, i.e. 0.68~\AA~, all four
bonds can be reasonably considered as actual hydrogen bonds. In difference from Str-d, the bond lengths in Str-h are distributed in a fairly wide range, from 1.70~\AA~to 3.64~\AA. It therefore leads to a higher total energy of Str-h compared with Str-d due to the absence of a bond, although the interaction
strength of the 1.70~\AA~ and 1.91~\AA~ bonds are rather strong. In
terms of Str-3, its bond lengths are very close to each
other, in the range from 2.12~\AA~ to 2.58~\AA. The quadrangle
formed by these four bonds is rather similar to a square, indicating a good
balance of hydrogen bonding strength in both directions.
It is still inconclusive which configuration is the one
that was observed in the experiments showing an anisotropic FEL
behavior, after atomistic structures of those configurations were
considered. It calls for a further investigation to distinguish
these configurations.

\begin{table}[tpb]
\caption{Bond lengths of the four hydrogen bonds, as marked in
Fig. \ref{fig:str}, of configurations Str-d, Str-h, and Str-3.}
\begin{center}
\begin{tabular*}{7.7cm}{ccccc}
  \hline\hline

\parbox{3cm}{Hydrogen bonds:} & \parbox{1.0cm}{1} & \parbox{1.00cm}{2} & \parbox{1.0cm}{3} & \parbox{1.0cm}{4}\\
\hline
Str-d (\AA) & ~~2.61 & 2.00 & 2.68 & 2.36 ~~\\
Str-h (\AA) & ~~2.35 & 1.91 & 1.70 & 3.64 ~~\\
Str-3 (\AA) & ~~2.24 & 2.36 & 2.12 & 2.58 ~~ \\

\hline\hline
\end{tabular*}
\end{center}
\label{tab:hbond}
\end{table}

\subsection{Band Structures}

\begin{figure*}[tpb]
\includegraphics[width=14.5cm]{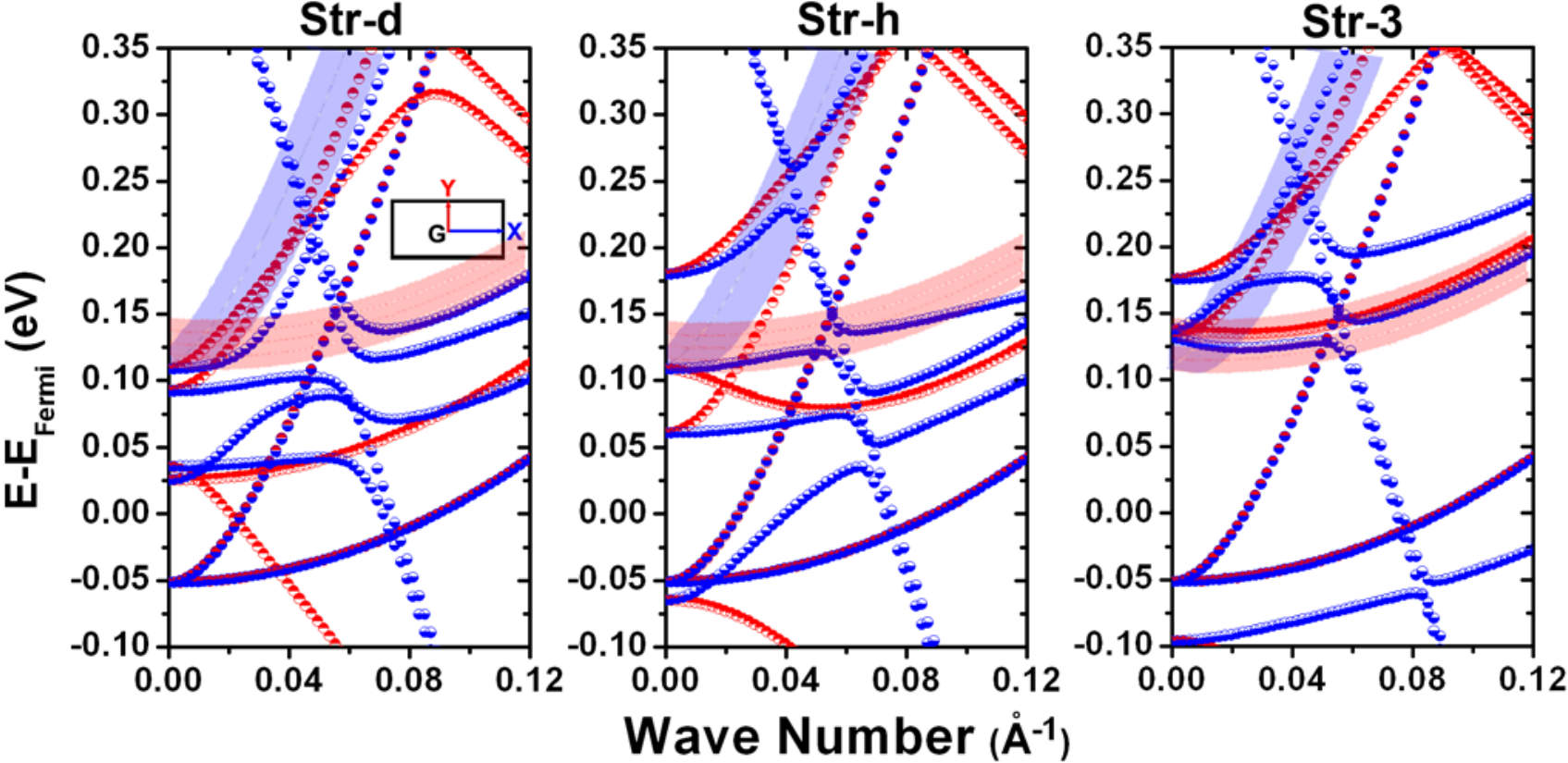}%
\caption{(Color online) Calculated band structures along [110] (blue
down-half-filled circle) and [$\overline{1}10$] (red up-half-filled
circle) of Str-d (a), Str-h (b), and Str-3 (c). Red and blue shaded
areas present the experimentally measured curves with a broadening
of 0.04~eV for the clarity of plotting. Both colors are consistent
throughout the paper, i.e. blue for [110] and red for
[$\overline{1}10$]. The first Brillouin zone together with three $k$
points (G, X, Y) were shown in (a) inset. } \label{fig:band}
\end{figure*}

We have calculated the band structures of configurations
Str-3, Str-d, and Str-h along the two directions from point G ($\Gamma$)
to X and Y, respectively, as shown in Fig. \ref{fig:band}. The reciprocal lattice vectors of the p$(2\times4)$ supercell is shown in the inset of Fig. \ref{fig:band} (left panel). The {\it k}-vector \overrightarrow{GX} corresponds to the [110] direction in real space (in blue for all panels) and \overrightarrow{GY} represents the [$\overline{1}10$] direction in real space (in red for all panels). Both colors are consistent with experimental plots, i.e. red for the [$\overline{1}10$] and blue for the [110] direction. In order to compare with the experimental measurement\cite{jacs2007}, the band-structure figures only show the dispersion relations from $\Gamma$ to the points away from $\Gamma$ by 0.2~\overrightarrow{GX} and 0.4~\overrightarrow{GY}, respectively (roughly 0.12
\AA$^{-1}$ in length). Red and blue shaded areas in Fig. \ref{fig:band} present the experimentally measured curves with a broadening of 0.04~eV. In the
experiment\cite{jacs2007}, a FEL state in the [$\overline{1}10$]
direction (in red, \overrightarrow{GY} in $k$-space), with an
electron effective mass $m_{e}$*$\sim$~0.6, was observed around 0.11
eV above the $E_{Fermi}$ at $\Gamma$. Another FEL state, starting from
the same energy at the $\Gamma$, with an $m_{e}$*$\sim$~0.06, was
detectable only in the [110] direction (\overrightarrow{GX} in
$k$-space, in blue).

The calculated band-structure of every configuration shows two
perfect parabolic bands starting from roughly -0.05 eV at the $\Gamma$,
which are observable in both directions (both in red and blue).
These two bands are assigned as surface states of the bare Cu
surface\cite{cu_ss} at the bottom of the slab. It differs within
0.01eV for the calculated values among the three molecular
configurations, indicating again that the energy resolution of our
calculation is better than 0.01~eV.

According to the band-structure of Str-d (left panel of Fig.
 \ref{fig:band}), except the state located at -0.05~eV, only one band
in red shows a FEL dispersion relation with $m_{e}$*$\sim$~0.6,
which resides 0.03~eV above the $E_{Fermi}$. Another band, with a
$m_{e}$*$\sim$~0.06, has almost the same eigenvalue at the $\Gamma$ and
reaches 0.35 eV at roughly 0.075 \AA$^{-1}$, although there is a
band-crossing induced gap opening from 0.05~eV to 0.13~eV (actual
gap from 0.09~eV to 0.11~eV). Even if the gap opening is not a
problem for STM-detection, the energy of these two bands at the
$\Gamma$ point, in between 0.03 eV and 0.04 eV, still cannot offer
a comparable result with the experimental observation. We therefore
can rule out configuration Str-d as that observed experimentally.

The result of Str-h, as shown in Fig. \ref{fig:band} (middle panel),
is even further away from the experimental result than that of
Str-d. Two bands hybridized by the Cu surface states and the
molecular states of glycine are located at 0.06~eV above
the $E_{Fermi}$. Apparently, it shows an anisotropic FEL behavior with a
very small gap opening in the direction shown in blue. However, the
effective masses of these two bands, i.e. smaller (larger) mass for
the red (blue) band, are opposite to the experiment (smaller mass for the blue). We therefore have to exclude Str-h as a likely candidate for the p($2\times4$) superstructure measured by STS.

Finally, the right panel of Fig. \ref{fig:band} shows the calculated
band structure of Str-3. It was found that two bands sitting around
0.13~eV at the $\Gamma$ point are capable of showing the anisotropic
FEL behavior. Both bands show an impressive consistency with the
experimental measurements, though slight gap openings appear in the
blue curves, i.e. at around 0.02~\AA$^{-1}$ (wave number) /
0.17~eV (energy) for the blue band (smaller $m_{e}$*). In
particular, the calculated eigenvalues at different $k$ points of
both bands match the measured data fairly well in both directions,
notwithstanding the theoretical band bottoms are 0.02~eV higher than
the experimental value. The theory-experiment agreement is
quite impressive because the small difference, i.e. 0.02 eV, almost reaches the energy resolution of STS measurements and DFT calculations.

\begin{figure*}[btp]
\includegraphics[width=14.5cm]{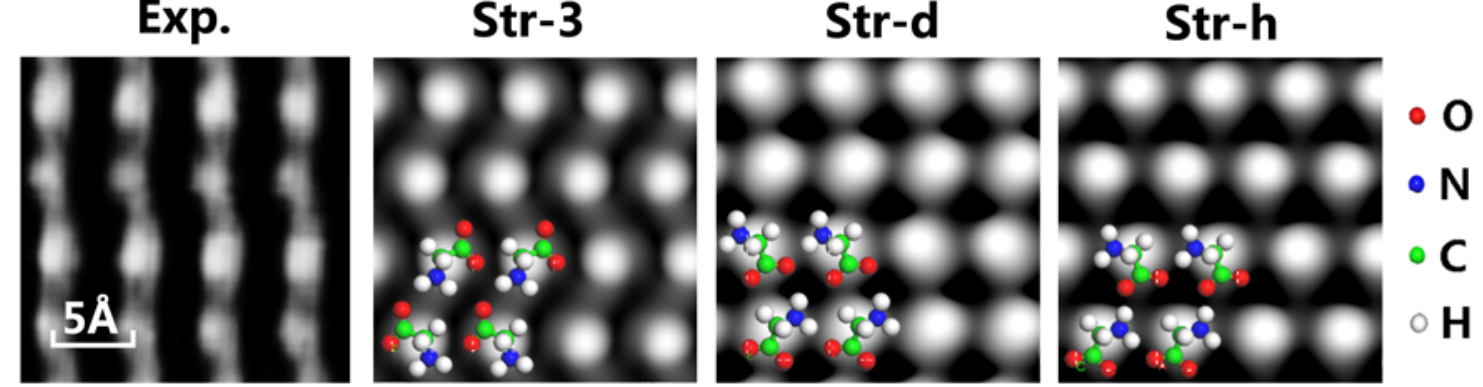}%
\caption{(Color online) Experimental STM image of the p($2\times4$) superstructure (Fig. 2(e) in Ref. \onlinecite{prl2007})(left) and simulated STM images for configurations Str-3 (middle left), Str-d(middle right) and Str-h(right), respectively. The sample bias in the simulation is -100 mV, the same as that in the experiment.} \label{fig:stm}
\end{figure*}

The electron effective mass of the two bands in the two directions
were fitted for configuration Str-3. In the red
direction ([$\overline{1}10$]), the fitted $m_{e}^{*}$ is 0.54,
which is 0.07 smaller than the measured value\cite{jacs2007} -
if all energy points in the band are used in the
fitting. In the experimental paper\cite{jacs2007}, however, the
$k$ value of the first data point is around 0.04 \AA$^{-1}$.
Therefore, we re-fit the theoretical data by removing energy points for
$k$ in between 0.00 and 0.04 \AA$^{-1}$ and the obtained $m_{e}^{*}$
is 0.60, which is only 0.01 smaller than the measured value of
0.61\cite{jacs2007}. In another direction, the fitted $m_{e}^{*}$
is 0.07, which only differs by 0.01 from the experimental value of
0.06\cite{jacs2007}. Notice that the fitted effective mass is
somewhat sensitive to the points near the $\Gamma$ point which were
not given in the experimental papers\cite{jacs2007,prl2007}. Our
results strongly suggest that configuration Str-3 is the one showing
the STS measured anisotropic FEL behavior.

\subsection{STM simulation}

The simulated STM images of configurations Str-3, Str-d, Str-h were compared with an available experiment. Figure \ref{fig:stm} shows an experimental STM image of the p($2\times4$) superstructure, together with the simulated STM images for Str-3, Str-d and  Str-h, respectively. The simulation was performed using the Tersoff-Hamann approximation. According to the experimental conditions, e.g. -100 mV sample bias\cite{prl2007}, the sample charge density was integrated from the states that their energy is in the range from -100 meV to the Fermi level (0meV). An iso-surface of charge density at 0.002e/\AA$^3$ is plotted in each panel.

Each simulated image contains several bright spots that are nearly centered around the upward H atoms connected with C atoms, as indicated by the atomic structure overlaid on the image. The image of Str-h shows a hexagonal structure, similar to the corresponding result reported in Ref. \onlinecite{DFT}, but which is completely different from that of the experiment. In terms of the position of the bright spots solely, the image of Str-d appears similar to that of Str-3. Both of them more or less reproduce the experimental observation and can hardly be distinguished. Notice that the experimental image also presents a ribbon-like feature, which is offered by Str-3 but not by Str-d. Although the STM image may be distorted by the subtle interaction between the tip and the upward H atom, the consistency of the ribbon-like feature between the experiment and the simulation of Str-3 suggests again that Str-3 is the most likely candidate for the p($2\times4$) superstructure observed in Ref. \onlinecite{prl2007}.

\section{FURTHER DISCUSSION and CONCLUSION}

There is an interesting issue that the STM/STS experiments are consistent with that of Str-3 but not with that of Str-d, although Str-d is energetically more favored than Str-3. Kinetic effect could most likely be responsible for the stabilization of configuration Str-3 against Str-d, according to a preliminary DFT calculation\cite{cu_step}. As a result, when the growth/evaporation rate is low, e.g. as that in molecular beam epitaxy, kinetic effect dominates. Therefore, Str-3
is kinetically favored, resulting in the formation of molecular
islands in the Str-3 configuration.

Notice that the atomic arrangement of N and O of Str-3 is the
same as that of Str-h. It means Str-3 also satisfies the PhD
experimental observations\cite{phd}. Our analysis presented above
suggests that the experimental preparation conditions in
Refs. \onlinecite{phd,jacs2007,prl2007} should
be similar, but they may substantially differ from that of
Refs. \onlinecite{yang99,yang01,yang02}, thereby leading to two observed
p(2$\times$4) superstructures, i.e. the kinetically favored Str-3 and the energetically favored Str-d.

In summary, according to various experimental and theoretical
results, the long-standing discrepancy regarding the exact
p(2$\times$4) structure of glycine on Cu(100) has been elucidated.
The exact configuration of the STS measured p(2$\times$4) superstructure is suggested Str-3, a new configuration not considered before, in
which the O-O line in a glycine is oriented roughly parallel to the
longer edge of the p(2$\times$4) supercell. This configuration has
chirality, satisfies the $<3\overline{1}0>$ rule, and also shows a relative good thermal stability. Its calculated band structure shows impressive agreements with the measured data, while other configurations cannot offer such agreements. Given the revealed configurations, interesting electronic behavior of this system can be further investigated. Finally, the Str-d configuration proposed in Ref. \onlinecite{yang99} and verified by DFT\cite{DFT}, is expected to be observable after an annealing treatment applied to a low-rate grown sample where Str-3 is predominant.

\section{Acknowledgments}
This work was financially supported by NSFC (11004244), BNSF
(2112019), FRFCU of MOE at RUC (10XNF085) (W.J.); and FQRNT of
Quebec, NSERC of Canada, and CIFAR (H.G.). Calculations were
performed at CLUMEQ in Canada and the PLHPC at RUC in China.

\end{document}